\documentclass[aps,prl,letterpaper,superscriptaddress,twocolumn,showpacs,floatfix,10pt]{revtex4-1}
\usepackage{amsfonts}
\usepackage{amsmath}
\usepackage{dsfont}
\usepackage{color}
\usepackage{amssymb}
\usepackage{graphicx}
\usepackage{epstopdf}
\usepackage{times}

\begin{document}
\title{Many-body localization in a disordered quantum Ising chain}
\author{Jonas A. Kj\"{a}ll}
\affiliation{Max-Planck-Institut f\"ur Physik komplexer Systeme, N\"othnitzer Str.\ 38, 01187 Dresden, Germany}
\author{Jens H. Bardarson}
\affiliation{Max-Planck-Institut f\"ur Physik komplexer Systeme, N\"othnitzer Str.\ 38, 01187 Dresden, Germany}
\author{Frank Pollmann}
\affiliation{Max-Planck-Institut f\"ur Physik komplexer Systeme, N\"othnitzer Str.\ 38, 01187 Dresden, Germany}

\begin{abstract}
Many-body localization occurs in isolated quantum systems when Anderson localization persists in the presence of finite interactions.
Despite strong evidence for the existence of a many-body localization transition a reliable extraction of the critical disorder strength is difficult due to a large drift with system size in the studied quantities.
In this work we explore two entanglement properties that are promising for the study of the many-body localization transition: the variance of the half-chain entanglement entropy of exact eigenstates and the long time change in entanglement after a local quench from an exact eigenstate. 
We investigate these quantities in a disordered quantum Ising chain and use them to estimate the critical disorder strength and its energy dependence.
In addition, we analyze a spin-glass transition at large disorder strength and provide evidence for it being a separate transition.
We thereby give numerical support for a recently proposed phase diagram of many-body localization with localization protected quantum order [Huse et al. Phys. Rev. B {\bf 88}, 014206 (2013)].
\end{abstract}

\maketitle
The Anderson insulator is an ideal insulator in which all single particle states are localized~\cite{Anderson:1958vr}.
This localization is due to quantum interference induced by elastic scattering of random impurities. 
In the absence of interactions and a coupling to a heat bath, the conductivity of such an insulator is identically zero at any temperature or energy. 
The fate of this insulating state when the particles interact with each other is the topic of the emerging field of {\it many-body localization} (MBL).
Following the seminal work of Basko et al.~\cite{Basko:2006hh}, it is generally believed that the insulating phase is stable to weak interactions.
This observation has fundamental consequences for our understanding and utilization of isolated quantum systems.
First, since a localized state cannot thermalize, it implies the breakdown of the eigenstate thermalization hypothesis~\cite{Deutsch:1991ju, Srednicki:1994dl,Rigol:2008bf} and thus the emergence of statistical physics from quantum mechanics.
Second, MBL could allow for the realization of topological and quantum order at finite energy densities by localizing the excitations that otherwise would melt the order~\cite{Huse:2013bw,Bauer:2013jw,Bahri:2013ux,Chandran:2013uz}.
The MBL phenomenon may also potentially be realized in multi-component systems, even in the absence of disorder~\cite{Grover:2013us,Schiulaz:2013to}, and in systems weakly coupled to heat baths~\cite{Nandkishore:vr}.

\begin{figure}[tbp]
  \begin{center}
    \includegraphics[width=80mm]{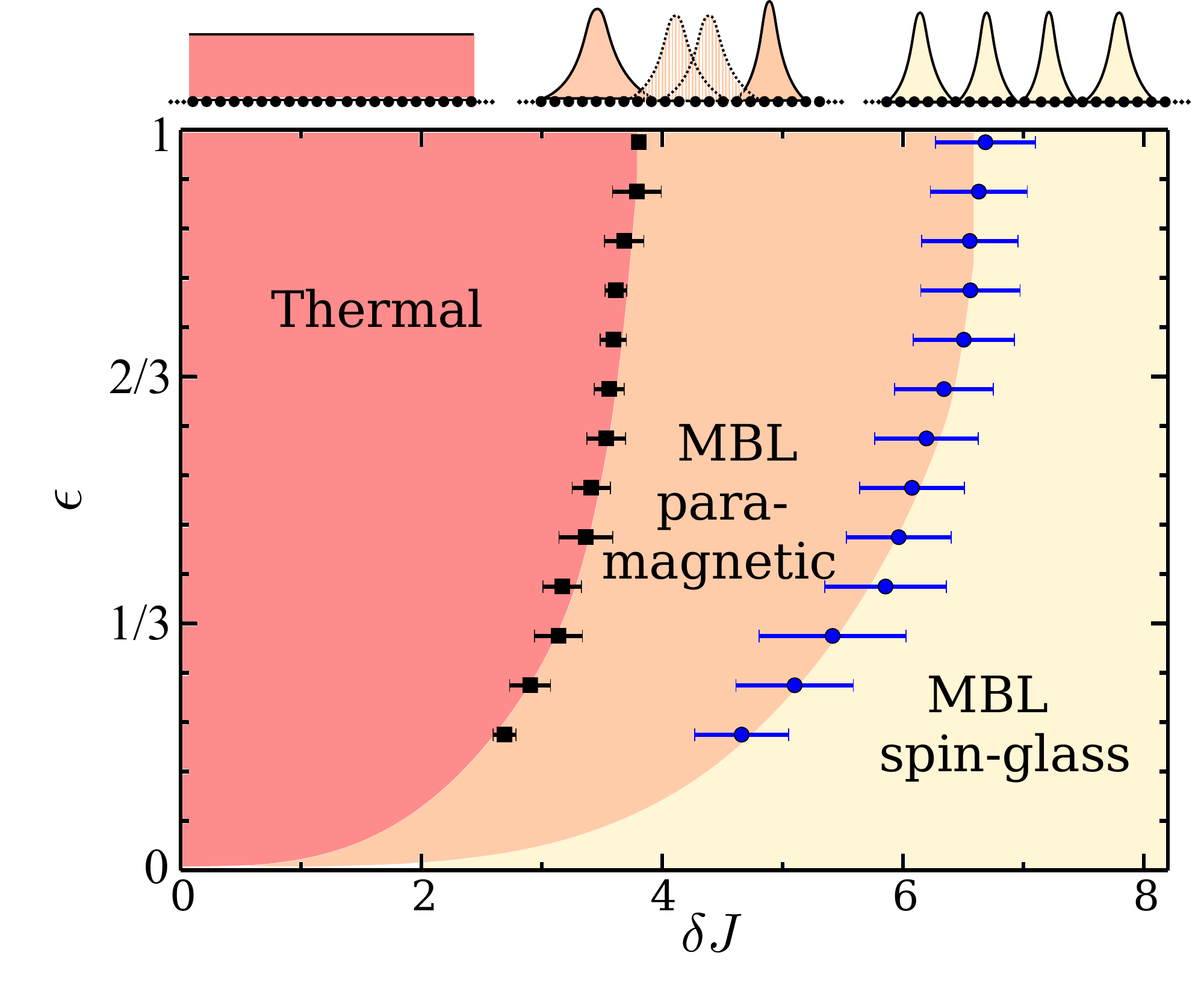}
    \caption{Phase diagram of the Ising model Eq.~\eqref{eq:H} in the $h/2=J_2=0.3$ plane with $\epsilon=2(E-E_{\mathrm{min}})/(E_{\mathrm{max}}-E_{\mathrm{min}})$ being the energy density relative to the total band width. The axes give the energy density above the ground state and the disorder strength. The colored areas are guides to the eye. The data is obtained from finite size scaling of entanglement difference Eq.~\eqref{eq:deltaS} after a local quench and the spin-glass order parameter Eq.~\eqref{eq:SG}; only statistical error bars are given, see text. The schematic on top of the phase diagram shows a caricature of the spatial domain wall probability distribution in the different phases. The thermal phase is characterized by extended domain walls, the MBL paramagnetic phase by localized domain walls which are created and removed in pairs (dashed), and the MBL spin-glass by localized non-overlapping domain walls.}
    \label{fig:PD}
  \end{center}
\end{figure}

A central concept in the phenomenology of MBL is that of the {\it many-body mobility edge}~\cite{Basko:2006hh}. 
Its existence implies that there is a definite energy at which the qualitative properties of the exact many-body eigenstates change: below that energy the eigenstates are close to a product state of localized single particle states; above, the eigenstates are extended and thermal, i.e., they follow the eigenstate thermalization hypothesis.
The presence of the many-body mobility edge does not contradict the identically zero conductivity of the many-body localized phase at nonzero temperature, as it would in the non-interacting case~\cite{Basko:2006hh}.
Rather, it suggests that the MBL transition is a dynamical transition and not an equilibrium phase transition.
In such a dynamical transition, the eigenstate properties change in a similar way as the ground state properties in conventional (quantum) phase transitions, as some control parameter, which for an MBL phase transition includes disorder strength and energy, is varied across its critical value~\cite{Huse:2013bw}.
Much insight into the nature of the many-body localized phase has been obtained from numerical studies of one-dimensional (1D) systems~\cite{Oganesyan:2007ex,Pal:2010gr,Monthus:2010gd,Berkelbach:2010ib,Barisic:2010ek,Igloi:2012in,Iyer:2013gy,DeLuca:2013ba,Pekker:2013vt,Vosk:2013ud} and in particular from studies of entanglement~\cite{Znidaric:2008cr,Bardarson:2012gc,Vosk:2013kt,Serbyn:2013he,Serbyn:2013cl,Bauer:2013jw,grover:2014}.
In the insulating phase, the exact eigenstates have relatively small entanglement and the entanglement entropy satisfies an area law, i.e., the von Neuman entropy of the reduced density matrix of a subsystem scales with the subsystem's surface area~\cite{Eisert:2010hq}. 
In the thermal phase the entanglement is extensive and satisfies a volume law.
Since the entanglement entropy is equivalent to the thermal entropy in a thermal state, its non-extensive nature in the localized phase reflects the absence of thermalization. 
In contrast, low entanglement states that are not eigenstates generically dephase logarithmically in time towards extensive but subthermal entanglement, even deep in the localized phase~\cite{Znidaric:2008cr,Bardarson:2012gc,Vosk:2013kt,Serbyn:2013he,Serbyn:2013cl}.
The drastically different entanglement properties of the localized and the thermal phase allow for an accurate determination of the MBL transition.
To clarify this statement, consider the half-chain entanglement entropy of an exact eigenstate of a 1D system at a fixed energy density.
In the vicinity of the MBL transition, small changes in either the energy density or the disorder realization can trigger a change from localized (area law) to a metallic state (volume law). 
The variance of the entanglement entropy over  an energy interval or disorder ensembles will therefore diverge with system size, signaling the MBL transition.
In a similar spirit, the MBL transition can be observed in the evolution of the entanglement entropy after a local quench from an exact eigenstate~\cite{Grover:2013us}.
That is, suppose we prepare the system in one of its eigenstates just below the mobility edge and then perturb it locally. 
The perturbation produces an uncertainty in the energy and the resulting state  is a linear combination of eigenstates  above and below the mobility edge.
This results in a diverging increase in the entanglement entropy as the state evolves from initially having an area law to eventually having a volume law entanglement.
In this work, we study the properties of all eigenstates of a 1D Ising chain model of MBL using full exact diagonalization~\cite{Sandvik:2010kc}.
In particular, through the study of entanglement, we detect the MBL transition and obtain an estimate of its energy dependence. 
In addition, similar to Ref.~\onlinecite{Pekker:2013vt}, we demonstrate the development of spin-glass (SG) order at large disorder strengths, with a transition that is separated from the localization transition.
The main results of our calculations are summarized in the phase diagram in Fig.~\ref{fig:PD} which agrees with that discussed qualitatively in a recent insightful work by Huse et al.~\cite{Huse:2013bw}.

We now turn to the details of our study. 
We employ the transverse field quantum Ising chain with disordered couplings and a next-nearest neighbor Ising term,
\begin{equation}
H=-\sum_{i=1}^{L-1} J_i\sigma_i^z\sigma_{i+1}^z + J_2\sum_{i=1}^{L-2}\sigma_i^z\sigma_{i+2}^z  +  h\sum_{i=1}^{L}\sigma_i^x,
\label{eq:H}
\end{equation}
where $\sigma^{x}$ and $\sigma^{z}$ are Pauli matrices and $L$ the number of sites in the chain.
The couplings $J_i=J+\delta J_i$ are random and independent, with all $\delta J_i$ taken from a uniform random distribution $[-\delta J,\delta J]$.
The  Hamiltonian~(\ref{eq:H}) has a global $\mathbb{Z}_2$ symmetry given by the parity operator $P = \prod_{i=1}^L\sigma_i^x$, with eigenvalues $\pm 1$.

When $J_2=\delta J = 0$, the model reduces to the well known quantum Ising chain in a transverse field.
A quantum critical point at $h=J$  separates a symmetry broken phase with ferromagnetic order ($h<J$) from a paramagnetic phase ($h>J$).
Since MBL is concerned with all energies, we are interested in the excited states, which in the ferromagnetic phase are (gapped) domain walls between different ferromagnetic domains. 
In the absence of disorder, the domain walls form extended states, with a dispersion proportional to $h$, and therefore destroy the order at any nonzero temperature (energy density above the ground state).
The model is one-dimensional, consequently any bond disorder ($\delta J > 0$) localizes the non-interacting domain wall excitations and the system forms an \emph{Anderson insulator}. 
The next-nearest neighbor coupling $J_2$ introduces a repulsive interaction between domain walls on adjacent bonds, and breaks the integrability of the model in the absence of disorder. 
In this work, we are primarily interested in the regime of  repulsive interactions in the ferromagnetic phase. 
For all the numerical results presented in this paper, we use the parameters $J=1$ and $h/2=J_2=0.3$. 
Our qualitative conclusions do not depend on the exact values of these parameters.

\begin{figure}[tbp]
  \begin{center}
    \includegraphics[width=90mm]{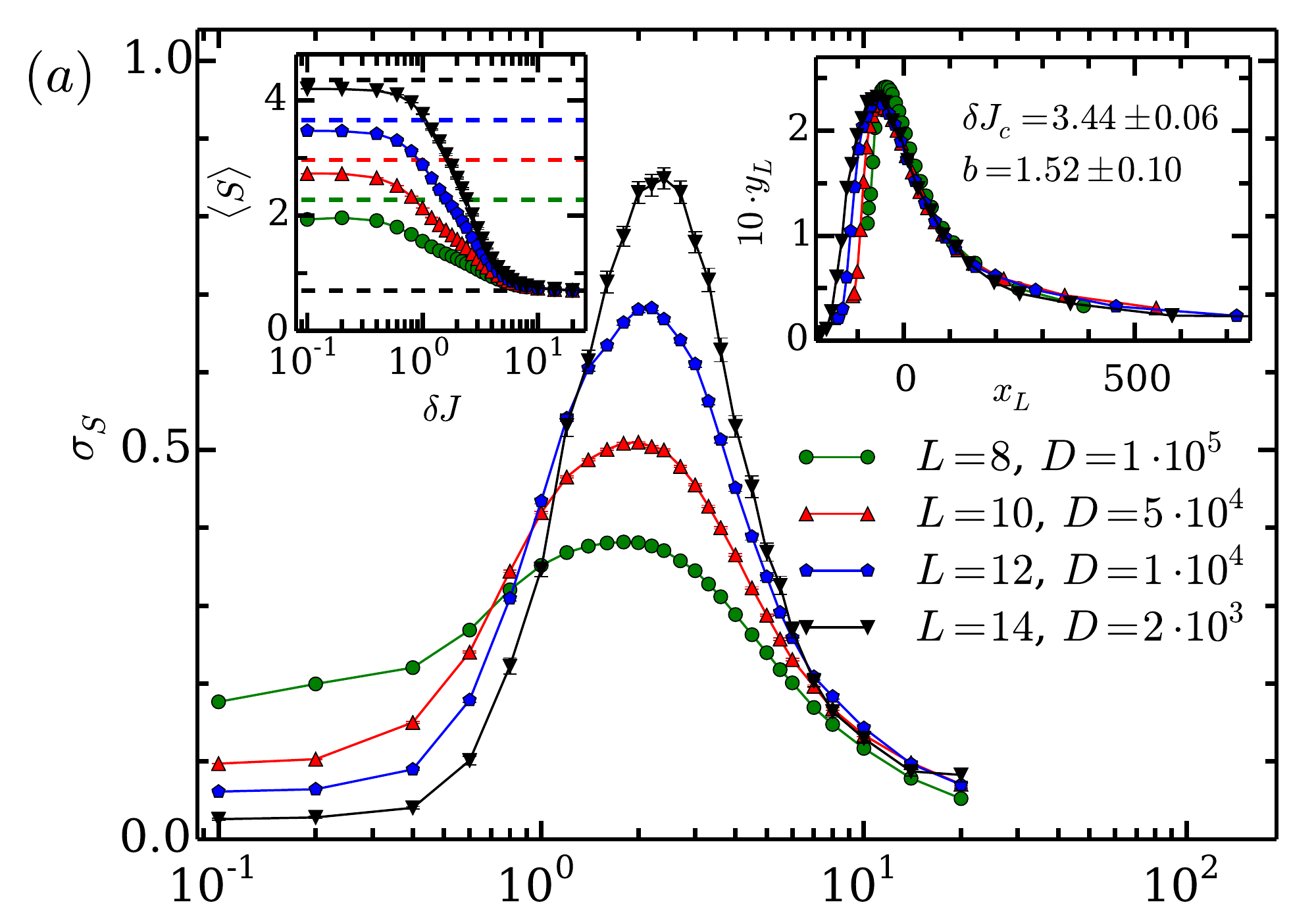}
    \includegraphics[width=90mm]{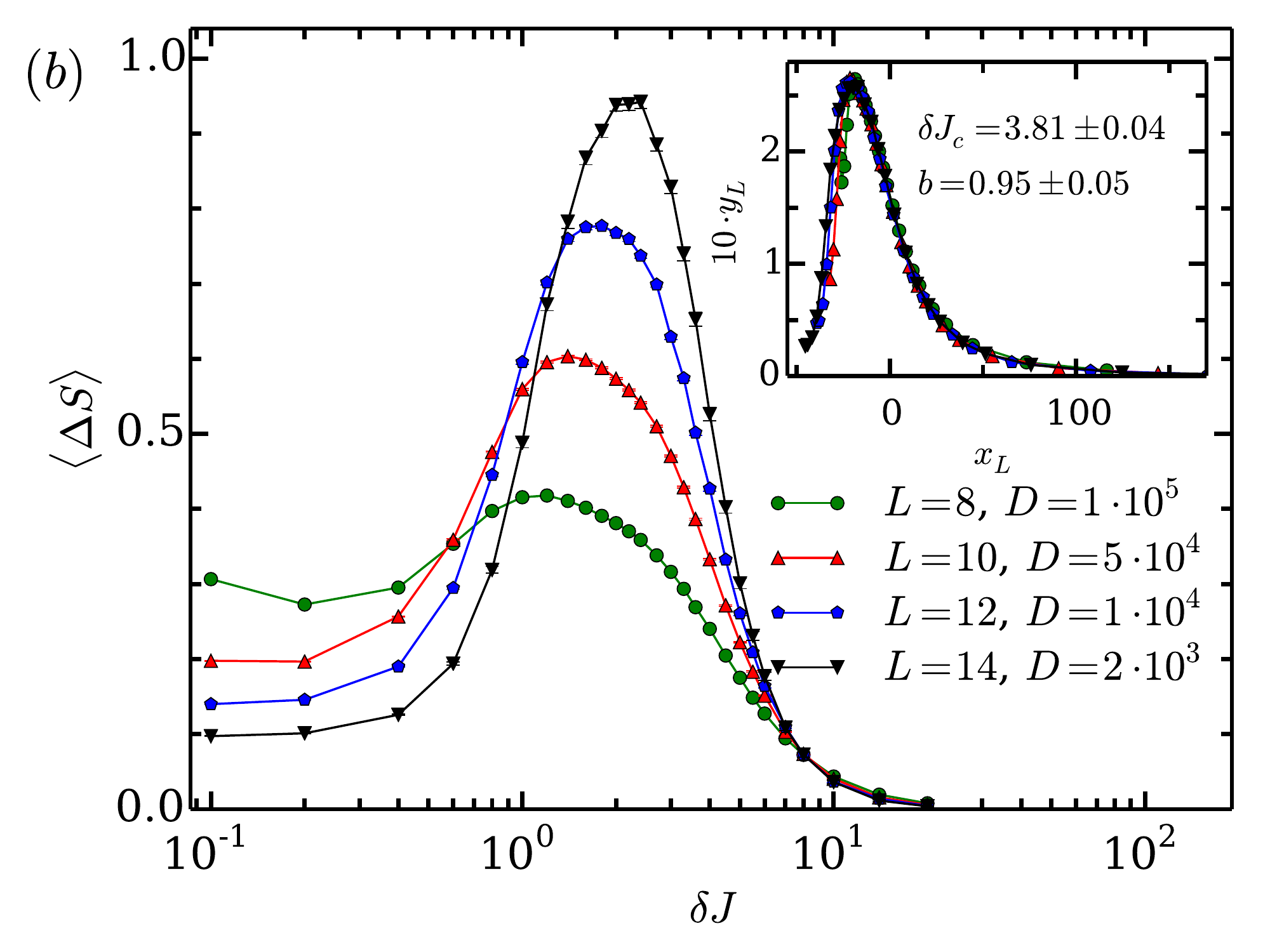}\\
    \caption{(a) Standard deviation of entanglement over the disorder ensemble as a function of disorder strength $\delta J$ for different system sizes $L$ and $D$ independent disorder realizations, at a fixed energy density in the middle of the spectrum ($\epsilon = 59/60$). The left inset show the mean entanglement entropy with dashed lines giving the values $S = (L\ln 2 - 1)/2$ and $S=\ln 2$. The right inset gives the scaling collapse of the data in the main panel. (b) Entanglement difference as a function of disorder strength for a local quench from an eigenstate in the middle of the spectrum. The inset gives the scaling collapse of the data.}
    \label{fig:DL}
  \end{center}
\end{figure}

At a fixed nonzero interaction strength a MBL transition is expected at a finite critical disorder strength $\delta J_c^\text{MBL}$, which generally depends on the energy density. 
An intuitive schematic picture of the nature of the different phases in terms of domain walls is given at the top of Fig~\ref{fig:PD}.
In the thermal phase domain walls are extended over the whole system, while in the MBL phase domain walls are localized.
Various approaches towards detecting the MBL transition have been adopted~\cite{Oganesyan:2007ex,Pal:2010gr,Bahri:2013ux,Monthus:2010gd,Berkelbach:2010ib}, but only a few have attempted a systematic finite size scaling analysis~\cite{Iyer:2013gy,Yao:2013vc,DeLuca:2013ba}, largely due to a significant drift of the studied quantities with system size.
We find the same problem with the level spacing statistics in the current model (data not shown) and therefore seek alternative quantities that allow for an accurate determination of phase boundaries.

We start by studying the entanglement in the exact eigenstates and focus on the half-chain entanglement entropy $S = -\text{Tr}_\text{L} \,\rho \ln \rho$ of the reduced density matrix $\rho = \text{Tr}_\text{R} |\psi\rangle\langle\psi|$, where the traces are over the left and right half-chain Hilbert spaces respectively.
For each disorder realization, we find the eigenstate $|n\rangle$ with energy $E_n$ closest to a fixed energy $E$ and thereby obtain a disorder distribution of entanglement entropies. 
In Fig.~\ref{fig:DL}a we plot the mean (left inset) and standard deviation of this distribution, at an energy in the middle of the spectrum, as a function of disorder strength.
In the thermal phase at weak disorder, the mean follows a volume law approaching the value $S = (L\ln 2 - 1)/2$ of a random state~\cite{Page:1993da} indicated by the dashed lines.
With increasing disorder the average entanglement entropy decreases and eventually saturates at $S = \ln 2$ deep in the localized phase.
The reason for this is that eigenstates become Schr\"odinger cat states with definite parity that are a linear combination of the two product states obtained from each other by the action of $P$, with each domain wall pinned by the disorder at a single bond. 
The standard deviation of the entanglement entropy goes to zero in the thermodynamic limit both deep in the thermal and localized phase, but diverges at the transition.
In the thermal phase this is consistent with the eigenstate thermalization hypothesis that requires the entropy to depend on energy only, while in the localized phase all states have the same $\ln 2$ entanglement entropy. 
The diverging peak could be understood as follows. For a given system size, disorder amplitude $\delta J$ and energy, near the transition $\delta J_c$, the exact value of the entanglement $S_n$ depends on the specific  disorder realization.
At a fixed value of $\delta J$ close to the transition, therefore, the set of states obtained from an ensemble of disorder realizations consists of both extended and localized states giving rise to a large standard deviation in the entanglement.
Naturally, with increasing system size the range of values of $\delta J$ that have states of mixed character narrows.
By the same token, we could observe the transition by measuring the standard deviation over small energy windows. 
Next we probe the MBL transition by studying the evolution of the entanglement entropy after a local quench at the edge of an eigenstate.
Before discussing the details of the physics, we explain the procedure we used.
After quenching an eigenstate $|n\rangle$ with a spin flip on the first site, we calculate the time dependent entanglement entropy $S_n(t)$ obtained from the von Neumann entropy of the state $|\psi_{n}(t)\rangle=\exp(-iHt)\sigma^x_1|n\rangle$.
In a finite system $S_n(t)$ saturates at long times allowing us to define the  difference of entanglement entropies
\begin{equation}
\Delta S_{n}=\lim_{t\rightarrow \infty} S_{n}(t)-S_{n}(0).
\label{eq:deltaS}
\end{equation}
In Fig.~\ref{fig:DL}b we plot the disorder averaged entanglement difference $\langle \Delta S \rangle$ as a function of disorder strength, at an energy in the middle of the spectrum.
The entanglement difference goes to zero both in the thermal and localized phases.
In the thermal phase, the entanglement difference goes to zero because of the eigenstate thermalization hypothesis since the local perturbation only introduces a small uncertainty in the energy of the state.
In the localized phase, the perturbation cannot propagate to the middle of the sample in order to generate any entanglement.
Note that the  perturbation of the exact eigenstate is local and therefore no entanglement is generated from the dephasing mechanism observed in a global quench~\cite{Bardarson:2012gc,Serbyn:2013he}. 

Around the transition, $\langle \Delta S \rangle$ peaks with a diverging amplitude. 
This diverging peak might be understood as a consequence of the many-body mobility edge. 
Namely, after the quench we have a state that is no longer an eigenstate, but rather a linear combination of a number of states with energies around $E_n$.
Close to the transition this linear combination contains both extended and localized state, and generates extensive entanglement under time evolution.
In the case when the initial state is a localized state, this results in an entanglement difference that scales with system size.
Unlike $\sigma_S$, the quench mixes eigenstates from the same disorder realization. 
Thus, a diverging $\langle \Delta S \rangle$ is suggestive for the existence of a many-body mobility edge.
The physics of this local quench might thus be related to that of the decay in Fock space of an electron-hole pair-excitation above an eigenstate as discussed by Basko et al.~\cite{Basko:2006hh}.

To determine the location of the phase transition, we perform a  scaling collapse separately on the standard deviation $\sigma_S$ and the entanglement difference $\langle \Delta S \rangle$, with a scaling function taking the form
\begin{equation}
Q(L,\delta J)=g(L)f[(\delta J-\delta J_c)L^{b}],
\label{eq:SCS}
\end{equation}
where $Q$ is the quantity that is scaled, $\delta J_c$ and $b$ are  scaling parameters, and $f$ is an undetermined function that is in principle different for the two quantities~\cite{Sandvik:2010kc}.
In the middle of the spectrum,  we expect the entanglement to fluctuate between the value $(L\log 2 -1)/2$ in the thermal phase and the value $\log 2$ of the strongly insulating phase.  
This motivates us to assume a prefactor of the form $g(L)= [(L-2)\log 2 - 1]/2$ near the transition point~\cite{PrivComm}.
At large system sizes this becomes a power law with unit exponent, compatible with the constraints on the entanglement entropy derived in Ref.~\onlinecite{grover:2014} , but the constant shifts are important to accurately collapse the data at the small system sizes available to us.
The collapsed data and the obtained scaling parameters are shown in the insets to Fig.~\ref{fig:DL}, where $x_L = (\delta J - \delta J_c)L^b$ and $y_L = Q/g(L)$.
Using this generic scaling function, we find a remarkably good collapse of the data.   
The given error bars on the critical disorder strength and exponents take into account only statistical errors, that is, they are obtained by repeating the scaling fit multiple times by adding noise to the data with amplitude given by the original error bars of the data~\cite{Sandvik:2010kc}. 
Since different scaling approaches (different choices for the amplitude $g$) give slightly different values for $\delta J_c$ and $b$, the actual error bars are considerably larger~\footnote{From exploring scaling collapses with different forms of $g(L)$, we believe that the values of $\delta J_c$ obtained from the $\Delta S$ and the $\sigma_S$ data agree within the actual error bars. It is currently unclear if the exponent $b$ should be the same for the two approaches (as one of them is dynamical), and if it should satisfy the bounds given by Chayes et al.~\cite{Chayes:1986kt}. For a more accurate scaling collapse, a better understanding of the underlying scaling theory (perhaps based on an infinite randomness fixed point~\cite{Pal:2010gr}) is needed.}.
At lower energy densities the entanglement in the thermal phase will be different from that of a random state, and the exact functional form of $g(L)$ is in this case therefore unknown.
Its form at high energies, however, suggests taking it as a sum of a term linear in $L$ and a constant (energy dependent) term.
Doing this, we obtain the energy dependence of the critical disorder strength $\delta J_c^{\text{MBL}}$ and thereby the phase boundary given in Fig.~\ref{fig:PD} (see the Appendix for details~\cite{SuppMat}).
We use the $\Delta S$ data as we find it to be slightly more accurate; the standard deviation gives consistent results.
The apparent phase boundary curvature hints at the presence of a many-body mobility edge~\cite{Basko:2006hh,Huse:2013bw}.

\begin{figure}[tbp]
  \begin{center}
    \includegraphics[width=90mm]{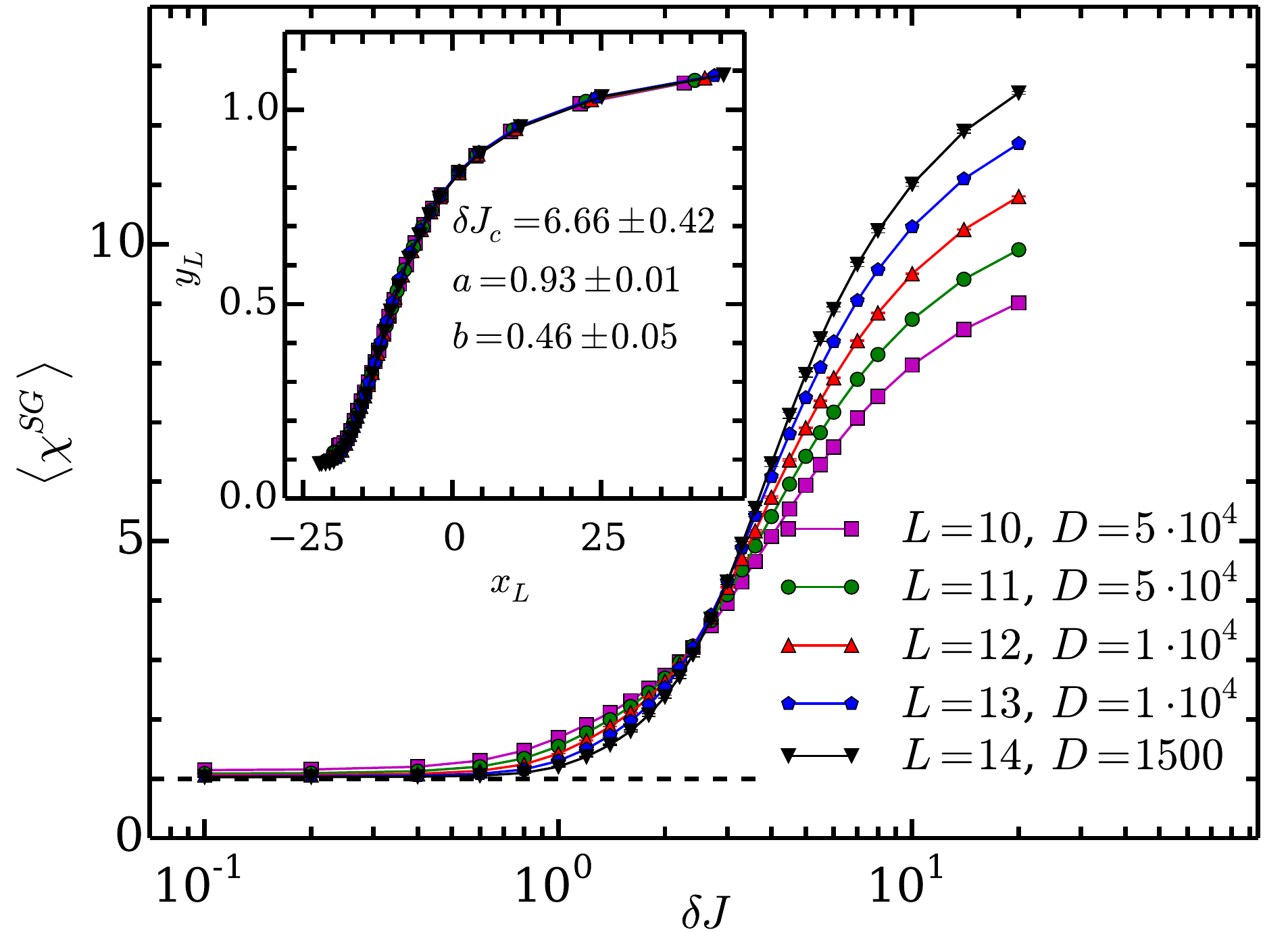}
    \caption{Spin-glass order parameter~\eqref{eq:SG} as function of disorder strength $\delta J$ for different system sizes $L$ and $D$ independent disorder realization, at a fixed energy in the middle of the energy spectrum ($\epsilon=59/60$). The dashed line gives the expected value $\langle \chi^\text{SG}\rangle = 1$, determined by normalization, in the absence of spin glass order. In the spin glass phase $\langle \chi^{\text{SG}}\rangle$ is proportional to $L$ . The inset shows the scaling collapse of the data.} 
    \label{fig:SGP}
  \end{center}
\end{figure}

The entanglement probes discussed above are only sensitive to the MBL transition, while as shown in Fig.~\ref{fig:PD} there are two separate MBL phases. 
If the number of domain walls is not conserved because overlapping pairs of domain walls are created and removed, the eigenstates have no order and form a paramagnet. 
A spin-glass order develops once the domain walls are strongly localized and their number fluctuations are small~\cite{Huse:2013bw,Pekker:2013vt}.
The regions between the separated domain walls have a fixed magnetization with the resulting broken $\mathbb{Z}_2$-symmetry protected by disorder.
The spin-glass order is reflected in the divergence of the order parameter
\begin{equation}
\label{eq:SG}
\chi^{\text{SG}}_{n}=\frac{1}{L}\displaystyle\sum_{i,j=1}^{L}\langle n| \sigma_i^z\sigma_j^z|n\rangle^2,
\end{equation}
in the thermodynamic limit ($\chi^\text{SG}\propto L$).
Outside the glassy phase $\chi^\text{SG} \rightarrow 1$ due to normalization.
The numerical results for $\langle\chi^{SG}\rangle$, obtained by averaging over all states within $58/60\leq\epsilon\leq1$ for each disorder realization, are shown in Fig.~\ref{fig:SGP}. 
As for the localization transition we obtain the spin-glass phase transition location by performing a finite size scaling via Eq.~(\ref{eq:SCS}) with $g(L)=L^a$ (see inset to Fig.~\ref{fig:SGP}).
Repeating this process for different energies results in the phase diagram given in Fig.~\ref{fig:PD}~\cite{SuppMat}.
From our data we conclude that the localization and spin-glass transition are separate transitions.

In addition to the dynamical MBL and spin-glass transitions a spectral transition takes place at large disorder strength~\cite{Huse:2013bw}, which unlike the other two transitions can not be directly detected in a single eigenstate.
Instead it is characterized by each energy in the spectrum becoming doubly degenerate. 
The reason for this degeneracy is the same as for the degenerate ground state of an open Ising chain: the presence of a Majorana edge mode bi-localized at the two edges.
In the spin glass phase the bulk excitations are localized strongly enough that the splitting of the Majorana mode is still exponentially small, going like $\exp(-L/\xi)$ with $\xi$ the localization length.
All states therefore come in pairs that  differ only in the occupation of the Majorana mode.
Unlike in the ground state, the broken $\mathbb{Z}_2$ symmetry is however not enough to give pairing, since the splitting of the edge modes competes with the mean level spacing, which is also exponentially small in system size but with a prefactor that depends only weakly on disorder.
The pairing transition could therefore be separate from the spin glass transition.
We have clearly observed the pairing transition in our data (not shown) but have not been able to perform a reliable finite size scaling analysis of it in order to obtain the transition location.
In particular, we could not determine with sufficient accuracy if the spectral transition is separate from the spin-glass transition, though our data suggests that it is.

In conclusion, we have explored two promising probes of the many-body localization transition and used them to study the transition in a disordered quantum Ising chain as a function of energy density. 
These probes are obtained from the entanglement properties of exact eigenstates, namely its standard deviation and its time evolution after a local quench at the edge of an eigenstate. 
We have obtained clear signatures of the many-body localization transition and gave evidence for the development of spin-glass order at large disorder strength.
Thereby, we provided a numerical estimate of the full MBL phase diagram as a function of disorder and energy that is consistent with that of Huse et al.~\cite{Huse:2013bw}.
We thank Miklos Gulacsi, Achilleas Lazarides, Arijeet Pal, and Tom Scheler for helpful discussions. We are very grateful for detailed discussions with Peter Young on spin-glass, and with Tarun Grover on entanglement following local quenches. We are especially grateful to David Huse and Arun Nanduri for critical reading of our manuscript and for providing numerous helpful suggestions that improved our draft.

\clearpage
\appendix

\section{Appendix: Energy dependence of the phase transitions and details about the statistical analysis performed}

In the appendix, we provide data for the energy density dependence of the localization and the spin-glass phase transition.
\begin{figure}[b]
  \begin{center}
    \includegraphics[width=79mm]{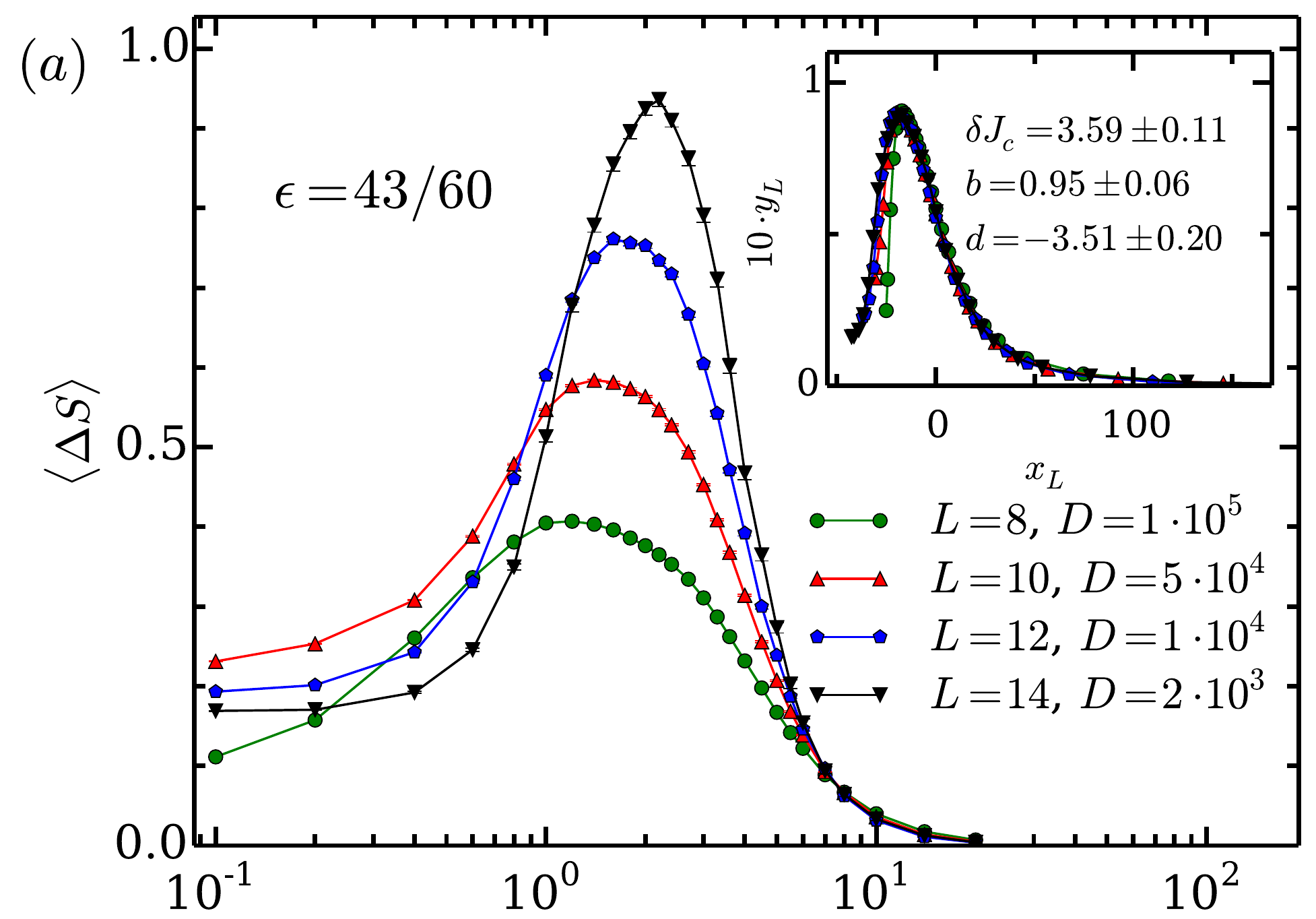}\\
            \includegraphics[width=79mm]{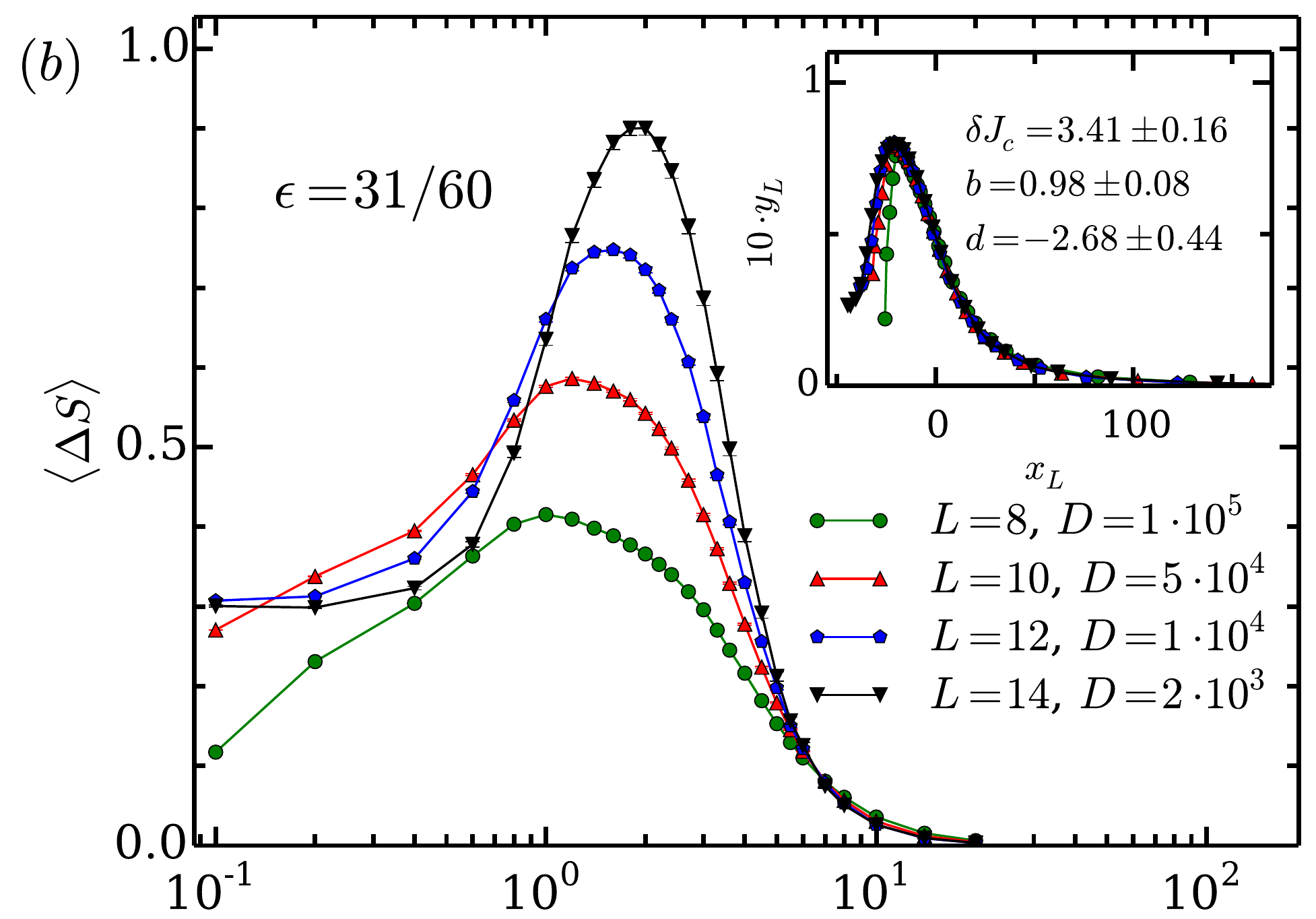}\\
                 \includegraphics[width=79mm]{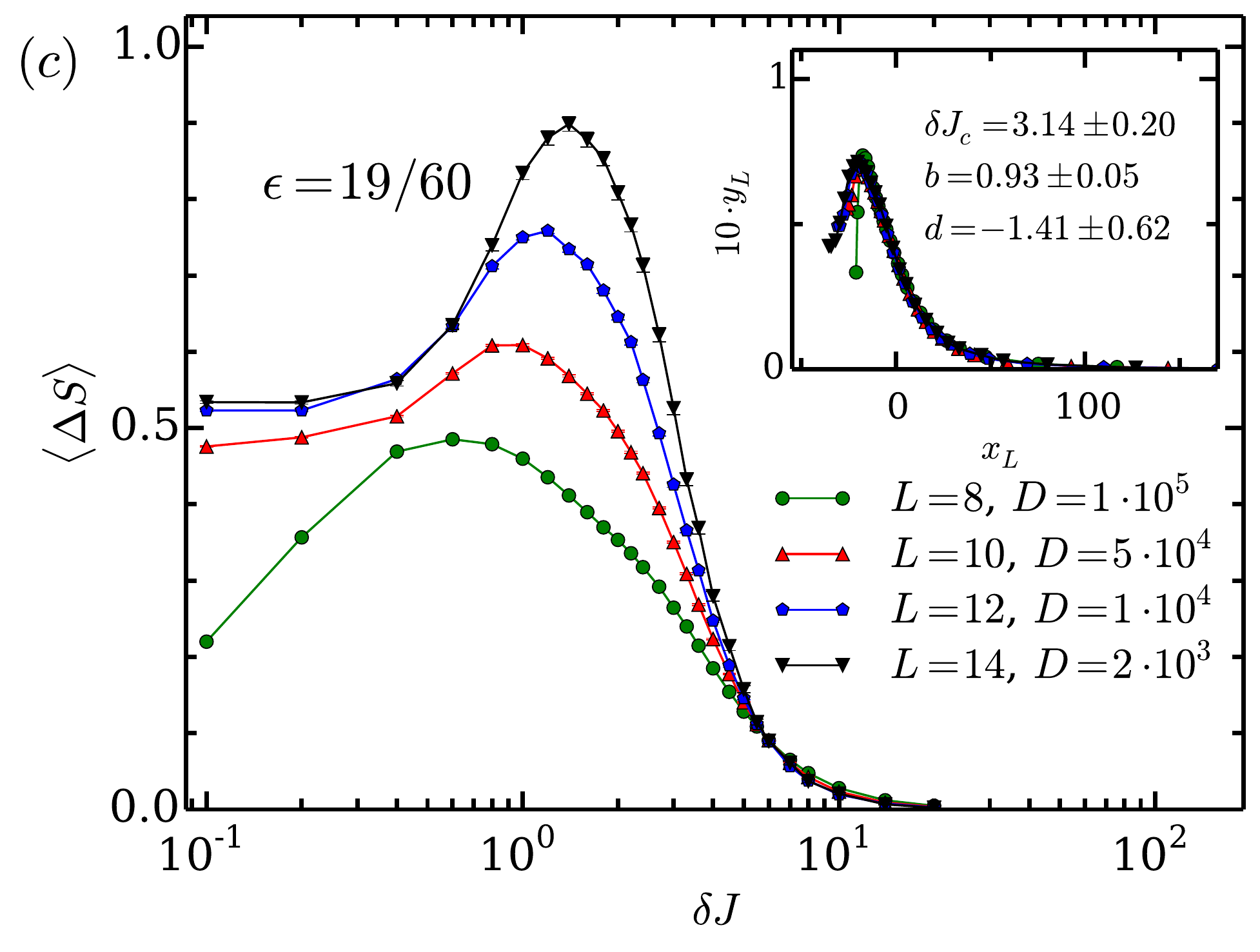}
    \caption{Entanglement entropy growth at a cut at bond $L/2$ after a local local quench at one of the edges for different energy densities, (a) $\epsilon=43/60$, (b) $\epsilon=31/60$, and (c) $\epsilon=19/60$. The insets show the scaling  collapse of the data yielding a critical disorder strength $\delta J_c^{\text{MBL}}$ and scaling parameters $b$ and $d$.}
    \label{fig:A1}
  \end{center}
\end{figure}
We also discuss the statistical analysis performed in this work and the error associated with our data.
In order to obtain the phase diagram shown in Fig.~1 in the main text, we repeated the analysis discussed in the main text for 15 equally spaced energy densities $\epsilon$ in the lower half of the energy spectrum.
We first show data for the energy density dependence of the entanglement entropy difference after a local quench $\langle\Delta S(\epsilon)\rangle$.
Examples of data, in addition to $\epsilon=59/60$ given in Fig.~2(b) of the main text, are shown in Fig.~\ref{fig:A1}.
\begin{figure}[tb]
  \begin{center}
    \includegraphics[width=79mm]{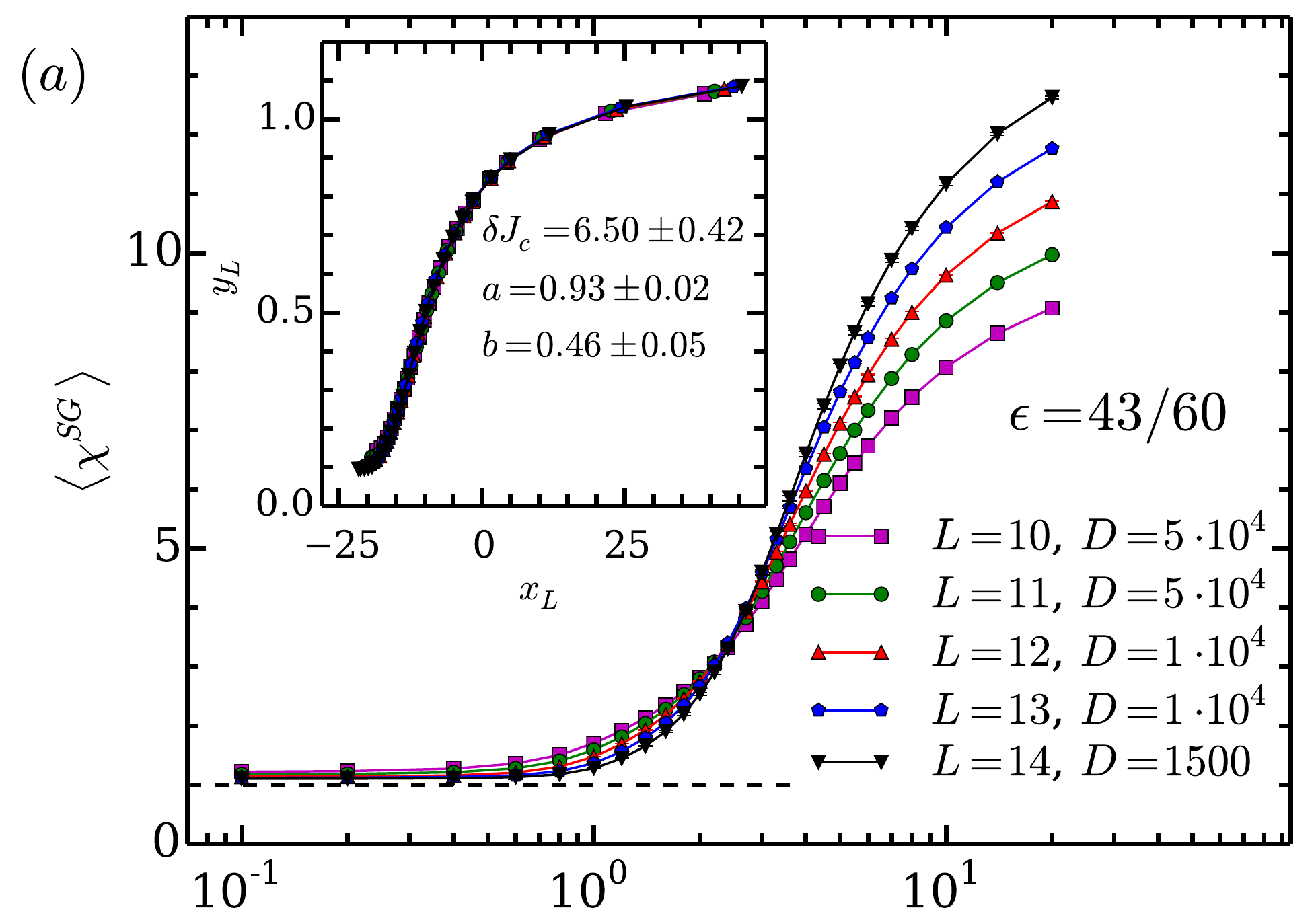}\\
            \includegraphics[width=79mm]{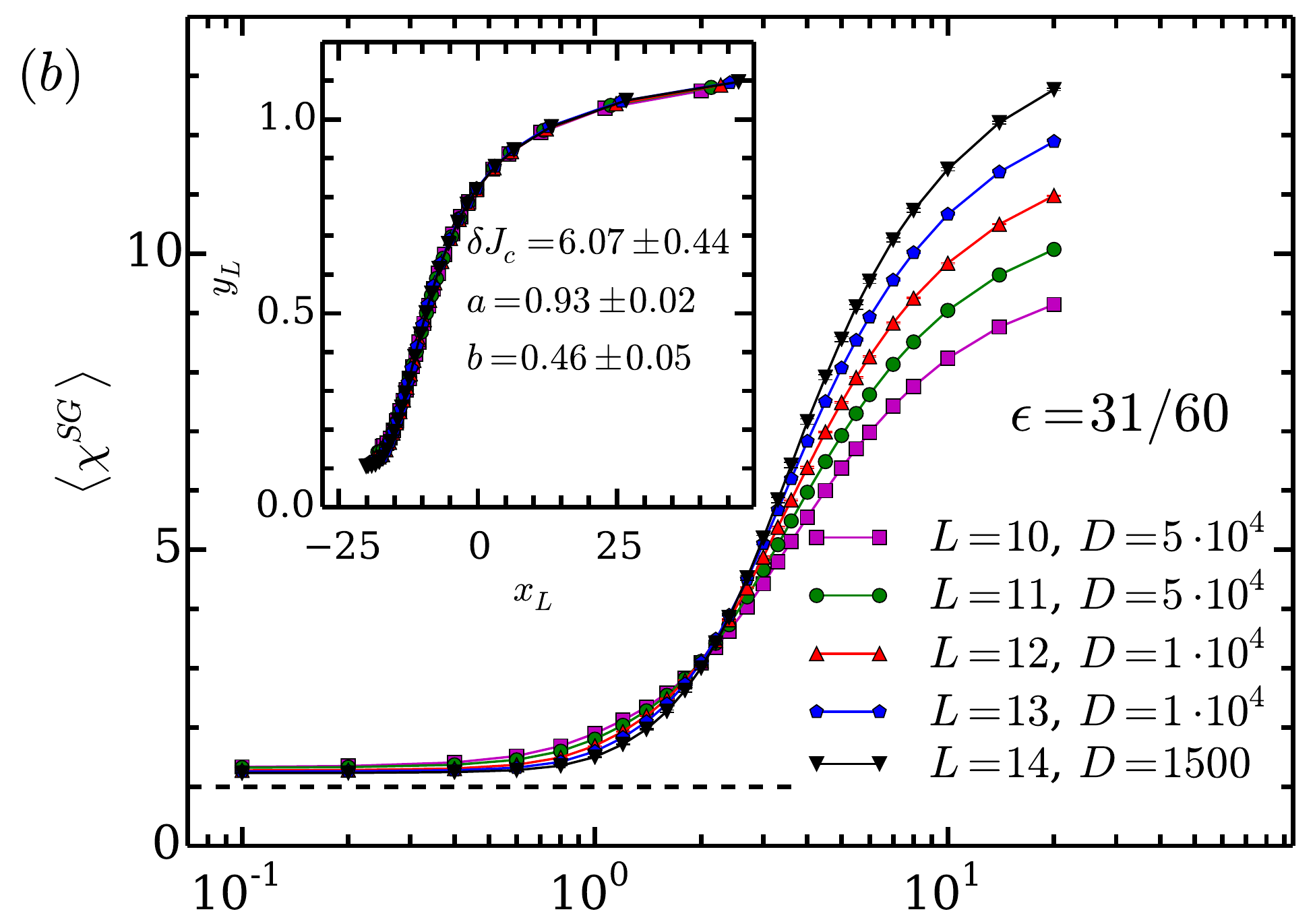}\\
                 \includegraphics[width=79mm]{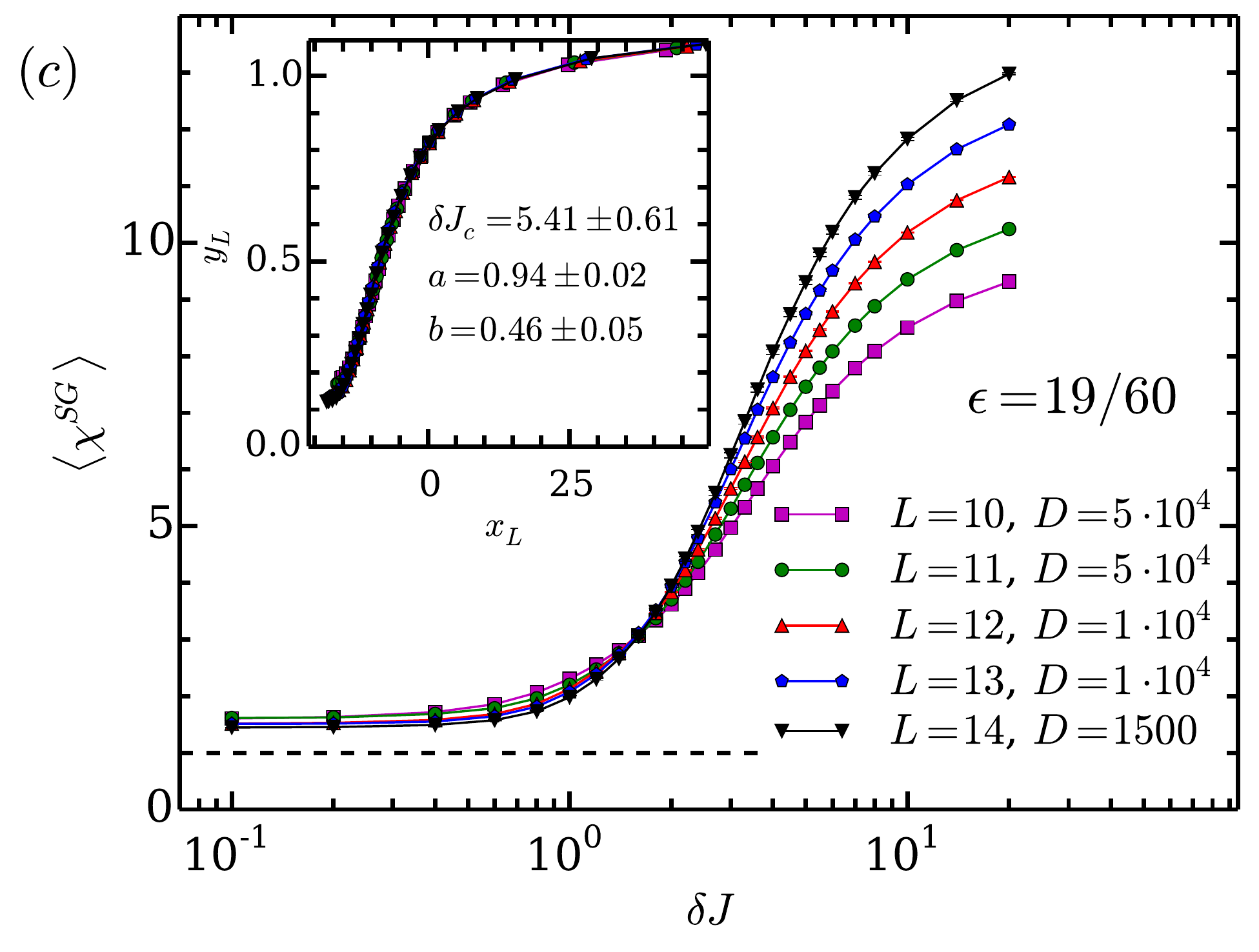}
    \caption{Spin-glass order parameter $\chi^\text{SG}$ as function of the disorder strength $\delta J$ for different energy densities, (a) $\epsilon=43/60$, (b) $\epsilon=31/60$, and (c) $\epsilon=19/60$. The insets show the scaling  collapse of the data with critical disorder strength $\delta J_c^{\text{SG}}$ and scaling parameters $a$ and $b$ (fixed).}

        \label{fig:A2}
  \end{center}
\end{figure}
To obtain the energy dependence of the entanglement properties, we take, for each disorder realization, the single eigenstate in the $P=-1$ sector closest to $\epsilon$ (apart from the case of $L=14$ where for $500$ disorder realizations, the average from the 20 eigenstates closest to the desired energy was used to decrease the errors).
Next, we present data for the spin-glass order parameter $\langle \chi^\text{SG}(\epsilon)\rangle$ as a function of energy density. 
Examples of data obtained for different energy densities are shown in the different panels of Fig.~\ref{fig:A2} (see also Fig.~3 of the main text for $\epsilon=59/60$).
For each disorder realization, the data are obtained from all eigenstates in the $P=-1$ parity sector, within an energy density range $\Delta \epsilon=1/30$ centered on $\epsilon$ for every disorder realization.

A scaling collapse of the finite size data is challenging since little is known about the critical behavior of the transitions we investigate, leaving us with many unknown parameters. 
Away from the middle of the spectra, the exact form of $g(L)$ [Eq.~(3) of the main text] is not known.
Instead, as argued in the main text, a three parameter scaling collapse with $g(L)=(1-dL^{-1})$ is used, see insets in  Fig.~\ref{fig:A1}.
Performing the same type of a three parameter scaling collapse in the middle of the spectra, gives consistent results with the two parameter scaling collapse used in the main paper, where $d = (1+2\log 2)/\log 2$ was fixed by physical considerations.
This supports our argument for the physical form of $g$ in the middle of the spectrum.
The scaling of the spin-glass order parameter $\langle \chi^\text{SG}(\epsilon)\rangle$ is done with two unknown parameters $a$ and $\delta J_c$ and fixed $b=0.46\pm 0.05$, the value obtained in Fig.~3 of the main text.
The error bars are obtained by repeating the scaling fit multiple times by adding noise to the data with amplitude given by the original error bars of the data~\cite{Sandvik:2010kc}.
For the lowest energies ($\epsilon\lesssim 7/60$), no reliable scaling collapse is possible and hence the critical values at these energies can not be obtained. 

We end the Appendix with a brief discussion of the statistical analysis performed in this paper. 
For each system size and disorder strength, the calculations are done for many different disorder realizations. 
For each of these realizations, where numerically feasible, we first average over the results obtained from nearby eigenstates (see details above for respective quantity) to decrease the variance in the data for the subsequent average and error bar calculations from the different disorder realizations.
Consequently, all error bars presented indicate the statistical error.
Other systematical errors, that can be severe, are not directly taken into account in the error bars.
A consistent way to check how statistically reasonable the analyzed data is, is to calculate the $\chi^2$-value of the data~\cite{Sandvik:2010kc}, which is a quantity obtained in the minimization of the scaling collapse.
Data without finite size effects and other systematical errors has $\chi^2\approx 1$ per degree of freedom.
We obtain somewhat larger $\chi^2$-values, indicating that the actual error bars should be larger.


\begin{thebibliography}{36}%
\makeatletter
\providecommand \@ifxundefined [1]{%
 \@ifx{#1\undefined}
}%
\providecommand \@ifnum [1]{%
 \ifnum #1\expandafter \@firstoftwo
 \else \expandafter \@secondoftwo
 \fi
}%
\providecommand \@ifx [1]{%
 \ifx #1\expandafter \@firstoftwo
 \else \expandafter \@secondoftwo
 \fi
}%
\providecommand \natexlab [1]{#1}%
\providecommand \enquote  [1]{``#1''}%
\providecommand \bibnamefont  [1]{#1}%
\providecommand \bibfnamefont [1]{#1}%
\providecommand \citenamefont [1]{#1}%
\providecommand \href@noop [0]{\@secondoftwo}%
\providecommand \href [0]{\begingroup \@sanitize@url \@href}%
\providecommand \@href[1]{\@@startlink{#1}\@@href}%
\providecommand \@@href[1]{\endgroup#1\@@endlink}%
\providecommand \@sanitize@url [0]{\catcode `\\12\catcode `\$12\catcode
  `\&12\catcode `\#12\catcode `\^12\catcode `\_12\catcode `\%12\relax}%
\providecommand \@@startlink[1]{}%
\providecommand \@@endlink[0]{}%
\providecommand \url  [0]{\begingroup\@sanitize@url \@url }%
\providecommand \@url [1]{\endgroup\@href {#1}{\urlprefix }}%
\providecommand \urlprefix  [0]{URL }%
\providecommand \Eprint [0]{\href }%
\providecommand \doibase [0]{http://dx.doi.org/}%
\providecommand \selectlanguage [0]{\@gobble}%
\providecommand \bibinfo  [0]{\@secondoftwo}%
\providecommand \bibfield  [0]{\@secondoftwo}%
\providecommand \translation [1]{[#1]}%
\providecommand \BibitemOpen [0]{}%
\providecommand \bibitemStop [0]{}%
\providecommand \bibitemNoStop [0]{.\EOS\space}%
\providecommand \EOS [0]{\spacefactor3000\relax}%
\providecommand \BibitemShut  [1]{\csname bibitem#1\endcsname}%
\let\auto@bib@innerbib\@empty
\bibitem [{\citenamefont {Anderson}(1958)}]{Anderson:1958vr}%
  \BibitemOpen
  \bibfield  {author} {\bibinfo {author} {\bibfnamefont {P.~W.}\ \bibnamefont
  {Anderson}},\ }\href@noop {} {\bibfield  {journal} {\bibinfo  {journal}
  {Phys. Rev.}\ }\textbf {\bibinfo {volume} {109}},\ \bibinfo {pages} {1492}
  (\bibinfo {year} {1958})}\BibitemShut {NoStop}%
\bibitem [{\citenamefont {Basko}\ \emph {et~al.}(2006)\citenamefont {Basko},
  \citenamefont {Aleiner},\ and\ \citenamefont {Altshuler}}]{Basko:2006hh}%
  \BibitemOpen
  \bibfield  {author} {\bibinfo {author} {\bibfnamefont {D.~M.}\ \bibnamefont
  {Basko}}, \bibinfo {author} {\bibfnamefont {I.~L.}\ \bibnamefont {Aleiner}},
  \ and\ \bibinfo {author} {\bibfnamefont {B.~L.}\ \bibnamefont {Altshuler}},\
  }\href@noop {} {\bibfield  {journal} {\bibinfo  {journal} {Ann. Phys.}\
  }\textbf {\bibinfo {volume} {321}},\ \bibinfo {pages} {1126} (\bibinfo {year}
  {2006})}\BibitemShut {NoStop}%
\bibitem [{\citenamefont {Deutsch}(1991)}]{Deutsch:1991ju}%
  \BibitemOpen
  \bibfield  {author} {\bibinfo {author} {\bibfnamefont {J.}~\bibnamefont
  {Deutsch}},\ }\href@noop {} {\bibfield  {journal} {\bibinfo  {journal} {Phys.
  Rev. A}\ }\textbf {\bibinfo {volume} {43}},\ \bibinfo {pages} {2046}
  (\bibinfo {year} {1991})}\BibitemShut {NoStop}%
\bibitem [{\citenamefont {Srednicki}(1994)}]{Srednicki:1994dl}%
  \BibitemOpen
  \bibfield  {author} {\bibinfo {author} {\bibfnamefont {M.}~\bibnamefont
  {Srednicki}},\ }\href@noop {} {\bibfield  {journal} {\bibinfo  {journal}
  {Phys. Rev. E}\ }\textbf {\bibinfo {volume} {50}},\ \bibinfo {pages} {888}
  (\bibinfo {year} {1994})}\BibitemShut {NoStop}%
\bibitem [{\citenamefont {Rigol}\ \emph {et~al.}(2008)\citenamefont {Rigol},
  \citenamefont {Dunjko},\ and\ \citenamefont {Olshanii}}]{Rigol:2008bf}%
  \BibitemOpen
  \bibfield  {author} {\bibinfo {author} {\bibfnamefont {M.}~\bibnamefont
  {Rigol}}, \bibinfo {author} {\bibfnamefont {V.}~\bibnamefont {Dunjko}}, \
  and\ \bibinfo {author} {\bibfnamefont {M.}~\bibnamefont {Olshanii}},\
  }\href@noop {} {\bibfield  {journal} {\bibinfo  {journal} {Nature}\ }\textbf
  {\bibinfo {volume} {452}},\ \bibinfo {pages} {854} (\bibinfo {year}
  {2008})}\BibitemShut {NoStop}%
\bibitem [{\citenamefont {Huse}\ \emph {et~al.}(2013)\citenamefont {Huse},
  \citenamefont {Nandkishore}, \citenamefont {Oganesyan}, \citenamefont {Pal},\
  and\ \citenamefont {Sondhi}}]{Huse:2013bw}%
  \BibitemOpen
  \bibfield  {author} {\bibinfo {author} {\bibfnamefont {D.~A.}\ \bibnamefont
  {Huse}}, \bibinfo {author} {\bibfnamefont {R.}~\bibnamefont {Nandkishore}},
  \bibinfo {author} {\bibfnamefont {V.}~\bibnamefont {Oganesyan}}, \bibinfo
  {author} {\bibfnamefont {A.}~\bibnamefont {Pal}}, \ and\ \bibinfo {author}
  {\bibfnamefont {S.~L.}\ \bibnamefont {Sondhi}},\ }\href@noop {} {\bibfield
  {journal} {\bibinfo  {journal} {Phys. Rev. B}\ }\textbf {\bibinfo {volume}
  {88}},\ \bibinfo {pages} {014206} (\bibinfo {year} {2013})}\BibitemShut
  {NoStop}%
\bibitem [{\citenamefont {Bauer}\ and\ \citenamefont
  {Nayak}(2013)}]{Bauer:2013jw}%
  \BibitemOpen
  \bibfield  {author} {\bibinfo {author} {\bibfnamefont {B.}~\bibnamefont
  {Bauer}}\ and\ \bibinfo {author} {\bibfnamefont {C.}~\bibnamefont {Nayak}},\
  }\href@noop {} {\bibfield  {journal} {\bibinfo  {journal} {J. Stat. Mech.}\
  }\textbf {\bibinfo {volume} {P09005}} (\bibinfo {year} {2013})}\BibitemShut
  {NoStop}%
\bibitem [{\citenamefont {Bahri}\ \emph {et~al.}()\citenamefont {Bahri},
  \citenamefont {Vosk}, \citenamefont {Altman},\ and\ \citenamefont
  {Vishwanath}}]{Bahri:2013ux}%
  \BibitemOpen
  \bibfield  {author} {\bibinfo {author} {\bibfnamefont {Y.}~\bibnamefont
  {Bahri}}, \bibinfo {author} {\bibfnamefont {R.}~\bibnamefont {Vosk}},
  \bibinfo {author} {\bibfnamefont {E.}~\bibnamefont {Altman}}, \ and\ \bibinfo
  {author} {\bibfnamefont {A.}~\bibnamefont {Vishwanath}},\ }\href@noop {}
  {\bibinfo  {journal} {arXiv:1307.4092}\ }\BibitemShut {NoStop}%
\bibitem [{\citenamefont {Chandran}\ \emph {et~al.}(2014)\citenamefont
  {Chandran}, \citenamefont {Khemani}, \citenamefont {Laumann},\ and\
  \citenamefont {Sondhi}}]{Chandran:2013uz}%
  \BibitemOpen
\bibfield  {journal} {  }\bibfield  {author} {\bibinfo {author} {\bibfnamefont
  {A.}~\bibnamefont {Chandran}}, \bibinfo {author} {\bibfnamefont
  {V.}~\bibnamefont {Khemani}}, \bibinfo {author} {\bibfnamefont {C.~R.}\
  \bibnamefont {Laumann}}, \ and\ \bibinfo {author} {\bibfnamefont {S.~L.}\
  \bibnamefont {Sondhi}},\ }\href {\doibase 10.1103/PhysRevB.89.144201}
  {\bibfield  {journal} {\bibinfo  {journal} {Phys. Rev. B}\ }\textbf {\bibinfo
  {volume} {89}},\ \bibinfo {pages} {144201} (\bibinfo {year}
  {2014})}\BibitemShut {NoStop}%
\bibitem [{\citenamefont {Grover}\ and\ \citenamefont
  {Fisher}()}]{Grover:2013us}%
  \BibitemOpen
  \bibfield  {author} {\bibinfo {author} {\bibfnamefont {T.}~\bibnamefont
  {Grover}}\ and\ \bibinfo {author} {\bibfnamefont {M.~P.~A.}\ \bibnamefont
  {Fisher}},\ }\href@noop {} {\bibinfo  {journal} {arXiv:1307.2288v1}\
  }\BibitemShut {NoStop}%
\bibitem [{\citenamefont {Schiulaz}\ and\ \citenamefont
  {M{\"u}ller}(2014)}]{Schiulaz:2013to}%
  \BibitemOpen
\bibfield  {journal} {  }\bibfield  {author} {\bibinfo {author} {\bibfnamefont
  {M.}~\bibnamefont {Schiulaz}}\ and\ \bibinfo {author} {\bibfnamefont
  {M.}~\bibnamefont {M{\"u}ller}},\ }\href@noop {} {\bibfield  {journal}
  {\bibinfo  {journal} {AIP Conference Proceedings}\ }\textbf {\bibinfo
  {volume} {1610}} (\bibinfo {year} {2014})}\BibitemShut {NoStop}%
\bibitem [{\citenamefont {Nandkishore}\ \emph {et~al.}()\citenamefont
  {Nandkishore}, \citenamefont {Gopalakrishnan},\ and\ \citenamefont
  {Huse}}]{Nandkishore:vr}%
  \BibitemOpen
  \bibfield  {author} {\bibinfo {author} {\bibfnamefont {R.}~\bibnamefont
  {Nandkishore}}, \bibinfo {author} {\bibfnamefont {S.}~\bibnamefont
  {Gopalakrishnan}}, \ and\ \bibinfo {author} {\bibfnamefont {D.~A.}\
  \bibnamefont {Huse}},\ }\href@noop {} {\bibinfo  {journal} {arXiv:1402.5971}\
  }\BibitemShut {NoStop}%
\bibitem [{\citenamefont {Oganesyan}\ and\ \citenamefont
  {Huse}(2007)}]{Oganesyan:2007ex}%
  \BibitemOpen
\bibfield  {journal} {  }\bibfield  {author} {\bibinfo {author} {\bibfnamefont
  {V.}~\bibnamefont {Oganesyan}}\ and\ \bibinfo {author} {\bibfnamefont
  {D.}~\bibnamefont {Huse}},\ }\href@noop {} {\bibfield  {journal} {\bibinfo
  {journal} {Phys. Rev. B}\ }\textbf {\bibinfo {volume} {75}},\ \bibinfo
  {pages} {155111} (\bibinfo {year} {2007})}\BibitemShut {NoStop}%
\bibitem [{\citenamefont {Pal}\ and\ \citenamefont {Huse}(2010)}]{Pal:2010gr}%
  \BibitemOpen
  \bibfield  {author} {\bibinfo {author} {\bibfnamefont {A.}~\bibnamefont
  {Pal}}\ and\ \bibinfo {author} {\bibfnamefont {D.~A.}\ \bibnamefont {Huse}},\
  }\href@noop {} {\bibfield  {journal} {\bibinfo  {journal} {Phys. Rev. B}\
  }\textbf {\bibinfo {volume} {82}},\ \bibinfo {pages} {174411} (\bibinfo
  {year} {2010})}\BibitemShut {NoStop}%
\bibitem [{\citenamefont {Monthus}\ and\ \citenamefont
  {Garel}(2010)}]{Monthus:2010gd}%
  \BibitemOpen
  \bibfield  {author} {\bibinfo {author} {\bibfnamefont {C.}~\bibnamefont
  {Monthus}}\ and\ \bibinfo {author} {\bibfnamefont {T.}~\bibnamefont
  {Garel}},\ }\href@noop {} {\bibfield  {journal} {\bibinfo  {journal} {Phys.
  Rev. B}\ }\textbf {\bibinfo {volume} {81}},\ \bibinfo {pages} {134202}
  (\bibinfo {year} {2010})}\BibitemShut {NoStop}%
\bibitem [{\citenamefont {Berkelbach}\ and\ \citenamefont
  {Reichman}(2010)}]{Berkelbach:2010ib}%
  \BibitemOpen
  \bibfield  {author} {\bibinfo {author} {\bibfnamefont {T.~C.}\ \bibnamefont
  {Berkelbach}}\ and\ \bibinfo {author} {\bibfnamefont {D.~R.}\ \bibnamefont
  {Reichman}},\ }\href@noop {} {\bibfield  {journal} {\bibinfo  {journal}
  {Phys. Rev. B}\ }\textbf {\bibinfo {volume} {81}},\ \bibinfo {pages} {224429}
  (\bibinfo {year} {2010})}\BibitemShut {NoStop}%
\bibitem [{\citenamefont {Bari{\v s}i{\'c}}\ and\ \citenamefont {Prelov{\v
  s}ek}(2010)}]{Barisic:2010ek}%
  \BibitemOpen
  \bibfield  {author} {\bibinfo {author} {\bibfnamefont {O.~S.}\ \bibnamefont
  {Bari{\v s}i{\'c}}}\ and\ \bibinfo {author} {\bibfnamefont {P.}~\bibnamefont
  {Prelov{\v s}ek}},\ }\href@noop {} {\bibfield  {journal} {\bibinfo  {journal}
  {Phys. Rev. B}\ }\textbf {\bibinfo {volume} {82}},\ \bibinfo {pages} {161106}
  (\bibinfo {year} {2010})}\BibitemShut {NoStop}%
\bibitem [{\citenamefont {Igl{\'o}i}\ \emph {et~al.}(2012)\citenamefont
  {Igl{\'o}i}, \citenamefont {Szatm{\'a}ri},\ and\ \citenamefont
  {Lin}}]{Igloi:2012in}%
  \BibitemOpen
  \bibfield  {author} {\bibinfo {author} {\bibfnamefont {F.}~\bibnamefont
  {Igl{\'o}i}}, \bibinfo {author} {\bibfnamefont {Z.}~\bibnamefont
  {Szatm{\'a}ri}}, \ and\ \bibinfo {author} {\bibfnamefont {Y.-C.}\
  \bibnamefont {Lin}},\ }\href@noop {} {\bibfield  {journal} {\bibinfo
  {journal} {Phys. Rev. B}\ }\textbf {\bibinfo {volume} {85}},\ \bibinfo
  {pages} {094417} (\bibinfo {year} {2012})}\BibitemShut {NoStop}%
\bibitem [{\citenamefont {Iyer}\ \emph {et~al.}(2013)\citenamefont {Iyer},
  \citenamefont {Oganesyan}, \citenamefont {Refael},\ and\ \citenamefont
  {Huse}}]{Iyer:2013gy}%
  \BibitemOpen
  \bibfield  {author} {\bibinfo {author} {\bibfnamefont {S.}~\bibnamefont
  {Iyer}}, \bibinfo {author} {\bibfnamefont {V.}~\bibnamefont {Oganesyan}},
  \bibinfo {author} {\bibfnamefont {G.}~\bibnamefont {Refael}}, \ and\ \bibinfo
  {author} {\bibfnamefont {D.}~\bibnamefont {Huse}},\ }\href@noop {} {\bibfield
   {journal} {\bibinfo  {journal} {Phys. Rev. B}\ }\textbf {\bibinfo {volume}
  {87}},\ \bibinfo {pages} {134202} (\bibinfo {year} {2013})}\BibitemShut
  {NoStop}%
\bibitem [{\citenamefont {De~Luca}\ and\ \citenamefont
  {Scardicchio}(2013)}]{DeLuca:2013ba}%
  \BibitemOpen
  \bibfield  {author} {\bibinfo {author} {\bibfnamefont {A.}~\bibnamefont
  {De~Luca}}\ and\ \bibinfo {author} {\bibfnamefont {A.}~\bibnamefont
  {Scardicchio}},\ }\href@noop {} {\bibfield  {journal} {\bibinfo  {journal}
  {EPL}\ }\textbf {\bibinfo {volume} {101}},\ \bibinfo {pages} {37003}
  (\bibinfo {year} {2013})}\BibitemShut {NoStop}%
\bibitem [{\citenamefont {Pekker}\ \emph {et~al.}()\citenamefont {Pekker},
  \citenamefont {Refael}, \citenamefont {Altman}, \citenamefont {Demler},\ and\
  \citenamefont {Oganesyan}}]{Pekker:2013vt}%
  \BibitemOpen
  \bibfield  {author} {\bibinfo {author} {\bibfnamefont {D.}~\bibnamefont
  {Pekker}}, \bibinfo {author} {\bibfnamefont {G.}~\bibnamefont {Refael}},
  \bibinfo {author} {\bibfnamefont {E.}~\bibnamefont {Altman}}, \bibinfo
  {author} {\bibfnamefont {E.}~\bibnamefont {Demler}}, \ and\ \bibinfo {author}
  {\bibfnamefont {V.}~\bibnamefont {Oganesyan}},\ }\href@noop {} {\bibinfo
  {journal} {arXiv:13073253}\ }\BibitemShut {NoStop}%
\bibitem [{\citenamefont {Vosk}\ and\ \citenamefont
  {Altman}(2014)}]{Vosk:2013ud}%
  \BibitemOpen
\bibfield  {journal} {  }\bibfield  {author} {\bibinfo {author} {\bibfnamefont
  {R.}~\bibnamefont {Vosk}}\ and\ \bibinfo {author} {\bibfnamefont
  {E.}~\bibnamefont {Altman}},\ }\href {\doibase
  10.1103/PhysRevLett.112.217204} {\bibfield  {journal} {\bibinfo  {journal}
  {Phys. Rev. Lett.}\ }\textbf {\bibinfo {volume} {112}},\ \bibinfo {pages}
  {217204} (\bibinfo {year} {2014})}\BibitemShut {NoStop}%
\bibitem [{\citenamefont {{\v Z}nidari{\v c}}\ \emph
  {et~al.}(2008)\citenamefont {{\v Z}nidari{\v c}}, \citenamefont {Prosen},\
  and\ \citenamefont {Prelov{\v s}ek}}]{Znidaric:2008cr}%
  \BibitemOpen
  \bibfield  {author} {\bibinfo {author} {\bibfnamefont {M.}~\bibnamefont {{\v
  Z}nidari{\v c}}}, \bibinfo {author} {\bibfnamefont {T.}~\bibnamefont
  {Prosen}}, \ and\ \bibinfo {author} {\bibfnamefont {P.}~\bibnamefont
  {Prelov{\v s}ek}},\ }\href@noop {} {\bibfield  {journal} {\bibinfo  {journal}
  {Phys. Rev. B}\ }\textbf {\bibinfo {volume} {77}},\ \bibinfo {pages} {064426}
  (\bibinfo {year} {2008})}\BibitemShut {NoStop}%
\bibitem [{\citenamefont {Bardarson}\ \emph {et~al.}(2012)\citenamefont
  {Bardarson}, \citenamefont {Pollmann},\ and\ \citenamefont
  {Moore}}]{Bardarson:2012gc}%
  \BibitemOpen
  \bibfield  {author} {\bibinfo {author} {\bibfnamefont {J.~H.}\ \bibnamefont
  {Bardarson}}, \bibinfo {author} {\bibfnamefont {F.}~\bibnamefont {Pollmann}},
  \ and\ \bibinfo {author} {\bibfnamefont {J.~E.}\ \bibnamefont {Moore}},\
  }\href@noop {} {\bibfield  {journal} {\bibinfo  {journal} {Phys. Rev. Lett.}\
  }\textbf {\bibinfo {volume} {109}},\ \bibinfo {pages} {017202} (\bibinfo
  {year} {2012})}\BibitemShut {NoStop}%
\bibitem [{\citenamefont {Vosk}\ and\ \citenamefont
  {Altman}(2013)}]{Vosk:2013kt}%
  \BibitemOpen
  \bibfield  {author} {\bibinfo {author} {\bibfnamefont {R.}~\bibnamefont
  {Vosk}}\ and\ \bibinfo {author} {\bibfnamefont {E.}~\bibnamefont {Altman}},\
  }\href@noop {} {\bibfield  {journal} {\bibinfo  {journal} {Phys. Rev. Lett.}\
  }\textbf {\bibinfo {volume} {110}},\ \bibinfo {pages} {067204} (\bibinfo
  {year} {2013})}\BibitemShut {NoStop}%
\bibitem [{\citenamefont {Serbyn}\ \emph
  {et~al.}(2013{\natexlab{a}})\citenamefont {Serbyn}, \citenamefont
  {Papi{\'c}},\ and\ \citenamefont {Abanin}}]{Serbyn:2013he}%
  \BibitemOpen
  \bibfield  {author} {\bibinfo {author} {\bibfnamefont {M.}~\bibnamefont
  {Serbyn}}, \bibinfo {author} {\bibfnamefont {Z.}~\bibnamefont {Papi{\'c}}}, \
  and\ \bibinfo {author} {\bibfnamefont {D.~A.}\ \bibnamefont {Abanin}},\
  }\href@noop {} {\bibfield  {journal} {\bibinfo  {journal} {Phys. Rev. Lett.}\
  }\textbf {\bibinfo {volume} {110}},\ \bibinfo {pages} {260601} (\bibinfo
  {year} {2013}{\natexlab{a}})}\BibitemShut {NoStop}%
\bibitem [{\citenamefont {Serbyn}\ \emph
  {et~al.}(2013{\natexlab{b}})\citenamefont {Serbyn}, \citenamefont
  {Papi{\'c}},\ and\ \citenamefont {Abanin}}]{Serbyn:2013cl}%
  \BibitemOpen
  \bibfield  {author} {\bibinfo {author} {\bibfnamefont {M.}~\bibnamefont
  {Serbyn}}, \bibinfo {author} {\bibfnamefont {Z.}~\bibnamefont {Papi{\'c}}}, \
  and\ \bibinfo {author} {\bibfnamefont {D.~A.}\ \bibnamefont {Abanin}},\
  }\href@noop {} {\bibfield  {journal} {\bibinfo  {journal} {Phys. Rev. Lett.}\
  }\textbf {\bibinfo {volume} {111}},\ \bibinfo {pages} {127201} (\bibinfo
  {year} {2013}{\natexlab{b}})}\BibitemShut {NoStop}%
\bibitem [{\citenamefont {Grover}()}]{grover:2014}%
  \BibitemOpen
  \bibfield  {author} {\bibinfo {author} {\bibfnamefont {T.}~\bibnamefont
  {Grover}},\ }\href@noop {} {\bibinfo  {journal} {arXiv:1405.1471}\
  }\BibitemShut {NoStop}%
\bibitem [{\citenamefont {Eisert}\ \emph {et~al.}(2010)\citenamefont {Eisert},
  \citenamefont {Cramer},\ and\ \citenamefont {Plenio}}]{Eisert:2010hq}%
  \BibitemOpen
\bibfield  {journal} {  }\bibfield  {author} {\bibinfo {author} {\bibfnamefont
  {J.}~\bibnamefont {Eisert}}, \bibinfo {author} {\bibfnamefont
  {M.}~\bibnamefont {Cramer}}, \ and\ \bibinfo {author} {\bibfnamefont {M.~B.}\
  \bibnamefont {Plenio}},\ }\href@noop {} {\bibfield  {journal} {\bibinfo
  {journal} {Rev. Mod. Phys.}\ }\textbf {\bibinfo {volume} {82}},\ \bibinfo
  {pages} {277} (\bibinfo {year} {2010})}\BibitemShut {NoStop}%
\bibitem [{\citenamefont {Sandvik}(2010)}]{Sandvik:2010kc}%
  \BibitemOpen
  \bibfield  {author} {\bibinfo {author} {\bibfnamefont {A.~W.}\ \bibnamefont
  {Sandvik}},\ }\href@noop {} {\bibfield  {journal} {\bibinfo  {journal} {AIP
  Conf. Proc.}\ }\textbf {\bibinfo {volume} {1297}},\ \bibinfo {pages} {135}
  (\bibinfo {year} {2010})}\BibitemShut {NoStop}%
\bibitem [{\citenamefont {Yao}\ \emph {et~al.}()\citenamefont {Yao},
  \citenamefont {Laumann}, \citenamefont {Gopalakrishnan}, \citenamefont
  {Knap}, \citenamefont {Mueller}, \citenamefont {Demler},\ and\ \citenamefont
  {Lukin}}]{Yao:2013vc}%
  \BibitemOpen
  \bibfield  {author} {\bibinfo {author} {\bibfnamefont {N.~Y.}\ \bibnamefont
  {Yao}}, \bibinfo {author} {\bibfnamefont {C.~R.}\ \bibnamefont {Laumann}},
  \bibinfo {author} {\bibfnamefont {S.}~\bibnamefont {Gopalakrishnan}},
  \bibinfo {author} {\bibfnamefont {M.}~\bibnamefont {Knap}}, \bibinfo {author}
  {\bibfnamefont {M.}~\bibnamefont {Mueller}}, \bibinfo {author} {\bibfnamefont
  {E.~A.}\ \bibnamefont {Demler}}, \ and\ \bibinfo {author} {\bibfnamefont
  {M.~D.}\ \bibnamefont {Lukin}},\ }\href@noop {} {\bibinfo  {journal}
  {arXiv:1311.7151}\ }\BibitemShut {NoStop}%
\bibitem [{\citenamefont {Page}(1993)}]{Page:1993da}%
  \BibitemOpen
\bibfield  {journal} {  }\bibfield  {author} {\bibinfo {author} {\bibfnamefont
  {D.}~\bibnamefont {Page}},\ }\href@noop {} {\bibfield  {journal} {\bibinfo
  {journal} {Phys. Rev. Lett.}\ }\textbf {\bibinfo {volume} {71}},\ \bibinfo
  {pages} {1291} (\bibinfo {year} {1993})}\BibitemShut {NoStop}%
\bibitem [{Pri()}]{PrivComm}%
  \BibitemOpen
  \href@noop {} {}\bibinfo {note} {Discussions with A.\ Nanduri and D.\ Huse on
  scaling properties at the MBL transition..}\BibitemShut {Stop}%
\bibitem [{Note1()}]{Note1}%
  \BibitemOpen
  \bibinfo {note} {From exploring scaling collapses with different forms of
  $g(L)$, we believe that the values of $\delta J_c$ obtained from the $\Delta
  S$ and the $\sigma _S$ data agree within the actual error bars. It is
  currently unclear if the exponent $b$ should be the same for the two
  approaches (as one of them is dynamical), and if it should satisfy the bounds
  given by Chayes et al.~\cite {Chayes:1986kt}. For a more accurate scaling
  collapse, a better understanding of the underlying scaling theory (perhaps
  based on an infinite randomness fixed point~\cite {Pal:2010gr}) is
  needed.}\BibitemShut {Stop}%
\bibitem [{Sup()}]{SuppMat}%
  \BibitemOpen
  \href@noop {} {}\bibinfo {note} {The raw data and the details of the scaling
  analysis are shown in the supplementary material.}\BibitemShut {Stop}%
\bibitem [{\citenamefont {Chayes}\ \emph {et~al.}(1986)\citenamefont {Chayes},
  \citenamefont {Chayes}, \citenamefont {Fisher},\ and\ \citenamefont
  {Spencer}}]{Chayes:1986kt}%
  \BibitemOpen
  \bibfield  {author} {\bibinfo {author} {\bibfnamefont {J.}~\bibnamefont
  {Chayes}}, \bibinfo {author} {\bibfnamefont {L.}~\bibnamefont {Chayes}},
  \bibinfo {author} {\bibfnamefont {D.}~\bibnamefont {Fisher}}, \ and\ \bibinfo
  {author} {\bibfnamefont {T.}~\bibnamefont {Spencer}},\ }\href@noop {}
  {\bibfield  {journal} {\bibinfo  {journal} {Phys. Rev. Lett.}\ }\textbf
  {\bibinfo {volume} {57}},\ \bibinfo {pages} {2999} (\bibinfo {year}
  {1986})}\BibitemShut {NoStop}%
\end{thebibliography}
\end{document}